\documentstyle[12pt]{article}
\begin{document}
\title{Optimal barrier subdivision for Kramers' escape rate}
\author{Mulugeta Bekele$^{1,}$\thanks{On leave from: Department of Physics,
Addis Ababa University, P.O. Box 1176, Addis Ababa, Ethiopia} , 
G. Ananthakrishna$^{2}$, and N. Kumar$^{3,1}$\\
$^1$Department of Physics and $^2$Materials Research Centre,\\
Indian Institute of Science, Bangalore 560 012, India\\
$^3$Raman Research Institue, C. V. Raman Avenue,\\
Bangalore 560 080, India}
\date{}
\maketitle
\noindent
{\bf Abstract}. We examine the effect of subdividing the potential 
barrier along the reaction coordinate on Kramers' escape rate for a 
model potential. Using the known supersymmetric potential approach, 
we show the existence of an optimal number of subdivisions 
that maximises the rate. 


\noindent
{\bf PACS Nos  05.70.L,31.70.H,87.15.R}

\noindent
{\bf 1. Introduction}

\noindent
The problem of surmounting a potential or, more generally, a free
energy barrier is a classical problem that appears in all processes
having thermally activated kinetics. This problem was originally
addressed by Kramers in the context of a bistable potential energy
curve [1]. He provided an approximate solution for the rate of escape
over the barrier in a high barrier low noise limit. In the commonly
encountered high friction limit, the bistable potential is usually
parameterized in terms of the height of the barrier at the potential
maximum and the width of, or the distance under, the barrier
connecting the initial and the final states (potential minima).
Since Kramers' original work, there has been a number of refinements
as well as varied novel applications of his solution, and a large
volume of literature exists on this [2,3].

There are, however, situations where the initial and the final states
are separated by a barrier which is so high that the estimated
reaction rate is very small, and yet the reaction actually turns out
to proceed at a substantially higher rate. The enhancement is
attributed to the catalytic action, notably of an enzyme in a
biochemical reaction that forms a `transition state' complex with the
substrate giving a reduced barrier height [4]. We, however, envisage
here an alternative scenario where the enzyme effectively reduces the
activation energy by subdividing the reaction path into a number of
discrete steps each requiring a much smaller barrier crossing. These
subdivisions are expected to correspond to the discrete
conformational/configurational changes of the macromolecules,
protiens say.  Besides looking for a physical consequence of the
barrier subdivision, in its own right, the problem can be viewed as
an excercise on rate processes in dissipative systems.  In the
present work we have considered the effect of the barrier subdivision
on the reaction rate in the high friction limit. This we have done
for a W-shaped model potential barrier whose subdivision can be well
parameterized. Our analysis of the problem is based on the
supersymmetric potential technique [5-7].

\noindent
{\bf 2. The Methodology}

\noindent
It is sufficient for our purpose to note that in the high friction
limit, Kramers' escape problem is one of solving the Smoluchowski
equation (SE):
\begin{equation}
\frac{\partial}{\partial  t}P(x,t) = D\, \frac{\partial}{\partial
\,x} \left(\frac{\partial}{\partial \, x} + \beta\, U^{\prime}
(x)\right)P(x,t),
\end{equation}
\noindent
where P(x,t) is the probability density associated with the particle
position, $U'(x)=\frac{dU}{dx}$ with $U(x)$ being the `double
well'-potential, $D$ is the diffusion constant and $\beta=(kT)^{-1}$
is the inverse temperature.  With the ansatz
\begin{equation}
P(x,t) = \phi (x)\, e^{-\beta U (x)/2} \, e^{-\lambda \, t}
\end{equation}
\noindent
the SE is converted to a Euclidean Schroedinger equation for $\phi$:
\begin{equation}
H_{+} \phi_{+}  = E_{+} \, \phi_{+}
\end{equation}
\noindent
with $H_{+} = A^{+}A$  being positive semi-definite, where $E_+=\lambda/D$
and,
\begin{equation}
A = \frac{\partial}{\partial \, x} + \frac{1}{2} \beta \, U^{\prime} (x),
\end{equation}
\begin{equation}
A^{+} = -\frac{\partial}{\partial x} + \frac{1}{2} \beta U^{\prime}(x).
\end{equation}
\noindent
This Hamiltonian $H_+$ corresponds to the motion of a particle in the
potential
\begin{equation}
V_{+} \, (x) = \left( \frac{1}{2} \beta \, U^{\prime} (x)\right)^{2}
-\frac{1}{2} \, \beta\, U^{\prime\prime}(x).
\end{equation}
\noindent
For a high barrier, the escape rate is determined by
the smallest nonzero eigenvalue, $\lambda_1=DE_+^1$, of the SE 
where $E_+^1$ is the eigenvalue of the first 
excited state of Eq.(3). On the other hand, this eigenstate is degenerate 
with the ground state $\phi_-^0$ of the `supersymmetric partner 
potential' $V_-(x)$ 
given by
\begin{equation}
V_{-} (x) = \left(\frac{1}{2} \, \beta \, U^{\prime} (x)
\right)^{2} + \frac{1}{2} \beta \, U^{\prime\prime} (x)
\end{equation}
\noindent
so that $H_-\,\phi_-^0=E_-\,\phi_-^0$ with $H_-=AA^+$ and $E_-=E_+^1$. 
The problem thus boils down to finding the ground state eigenvalue of 
this `partner' potential.

\noindent
{\bf 3. The model and its solution}

\noindent
3.1{\it The model potential and parameterization of subdivision}

\noindent
For the sake of simplicity we consider a symmetric W-potential. 
For a full characterisation of this potential we require two parameters, 
namely, the height $U_0$ and the width under the potential $2L_0$
(see Fig. 1a).  
We now subdivide the barrier between the initial and final states into a 
series of smaller connecting barriers of many steps (see Fig.1b). 
In order to examine the effect of barrier subdivision on the reaction
rate systematically, it is necessary to parameterize the subdivision
in a physically, reasonable manner. Consider the step 
located between $x_n$ and $x_{n+1}$ 
(Fig. 1d). We choose $U_1$, $U_2$ and the associated widths $a$,
$b$ (shown in the figure) such that $\frac{U_1}{a}=\frac{U_2}{b}$.
This choice simplifies the calculation further.
Note that $x_{n+1}-x_n=a+b$.
If we have a total of N such equally spaced steps from 
the top of the barrier on either side, then $L_0 = N a + (N-1) b$, 
while  $U_0 = N U_1 - (N-1) U_2$. We introduce a parameter 
$\rho$ defined as 
\begin{equation}
\rho \equiv\frac{(N-1)U_2}{N U_1}=\frac{(N-1) b}{N a}. 
\end{equation}

The aim is, given $U_0$ and $L_0$, to find the escape rate for
different values of barrier subdivision consistent with the high
barrier limit, i.e., for various values of N. Such parameterization
is physically reasonable as it not only keeps the barrier height and
its width fixed, it also keeps the area under the barrier
approximately constant as $N$ is varied.  Introducing a dimensionless
potential $u(x)=\frac{1}{2}\beta U(x)$, the `supersymmetric partner
potential' to $V_+(x)$, namely,$V_-(x)$, is then given by
\begin{equation}
V_-(x)=(u^{\prime}(x))^2  +  u^{\prime\prime}(x)
\end{equation}
\noindent
which for the considered potential takes the form
\begin{equation}
V_- (x)=\left(\frac{u_{1}}{a}\right)^{2} -
\left(\frac{2u_{1}}{a}\right)\,\sum\limits_{n=-N}^{+N} \,
\left [\, \delta (x-x_{n}) - \,\delta (x-x_{n}-a)\,\right]
\end{equation}
\noindent
where $u_1=\frac{1}{2}\beta U_1$. Note that the potential $V_-(x)$ is a
series of attractive and repulsive delta-potentials superimposed over
a constant potential (see Fig. 1c). Changing the variable $x$ to 
$y=\frac{x}{Na}$ leads to a new Hamiltonian $h_-$ given by
\begin{equation}
h_{-} \equiv a_{0}^{2} \, H_{-} = -\frac{\partial^{2}}{\partial y^{2}}
+ (u^{0}_{1})^{2} - 2u_{1}^{0}\, \sum\limits_{n=-N}^{+N}\, \left[\, \delta
(y - y_n) - \delta(y - y_n - a_1)\,\right].
\end{equation}
\noindent
where $a_0= Na$, $u_1^0= Nu_1$, $a_1= \frac{1}{N}$, 
$b_1= \frac{\rho}{N-1}$ and $y_n= n(a_1+b_1)$.   
With this, $h_{-} \phi_-^0(y) = e_{-}\, \phi_-^0(y)$, where $e_{-}$ 
is a dimensionless quantity equal to $a_0^2\,E_{-}$.

\noindent
3.2 {\it Solution}

\noindent
We use transfer matrix method to find the ground state energy. The ground 
state wave function $\phi_-^0$ is of the form $Ae^{-ky}+Be^{ky}$ peaked 
around the positions of the delta potentials. Consider one period of the 
potential, say, the interval between $y_n$ and $y_{n+1}$. Assume the wave 
function of the form
\begin{equation}
\phi_1(y) = A_n\,e^{-k(y-y_n)} \,+\, B_n\,e^{k(y-y_n)},
\end{equation}
\noindent
for the interval $y_{n-1}+a_1\le y \le y_n$ and the wave function of the
form
\begin{equation} 
\phi_2(y)= C_n\,e^{-k(y-y_n-a_1)}\,+\,D_n\,e^{k(y-y_n-a_1)}
\end{equation}
\noindent
for the interval $y_n \le y \le y_n + a_1$ with 
$k= [(u_1^0)^2 -e_-]^{\frac{1}{2}}$. By matching the wave function 
at $y_n$, i.e., $\phi_1(y_n)=\phi_2(y_n)$, and by integrating Eq.(11) 
around $y_n$ noting
that there is a negative delta-potential of strength $2u_1^0$, i.e., 
$\phi_1^{\prime}(y_n)\,-\,\phi_2^{\prime}(y_n)\,-2u_1^0\phi_1(y_n)\,=\,0$,
we get    
\begin{equation}
{C_{n}\choose D_{n}} = \pmatrix{
(1+\alpha)e^{-ka_1} & \alpha\,e^{-ka_1}  \cr
-\alpha\,e^{ka_1} & (1-\alpha)e^{ka_1}\cr}\,\, {A_{n}\choose
B_{n}}={\bf T_1}{A_n\choose B_n}
\end{equation}
relating the two pairs of amplitudes
($\alpha= \frac{u_1^0}{k}$). Next, matching the wave function having
amplitudes $C_n$ and $D_n$ with the wave function having amplitudes 
$A_{n+1}$ and $B_{n+1}$ (found in the interval $y_n+ a_1\le y \le y_{n+1}$)
and integrating Eq.(11) around $y_n+a_1$ noting that a positive 
delta-potential of strength $2u_1^0$ is located there, we get  
\begin{equation}
{A_{n+1}\choose B_{n+1}} = \pmatrix{(1-\alpha)e^{-kb_1} &
-\alpha\,e^{-kb_1} \cr
\alpha\,e^{kb_1} & (1+\alpha)e^{kb_1}\cr}\,\, {C_{n}\choose
D_{n}}={\bf T_2}{C_n\choose D_n}.
\end{equation}
\noindent
The transfer matrix, ${\bf T}$, relating the amplitudes 
$A_{n+1}$, $B_{n+1}$ 
to the amplitudes $A_{n}$ and $B_{n}$: 
\begin{equation}
{A_{n+1}\choose B_{n+1}} = {\bf T} {A_{n}\choose B_{n}}
\end{equation}
will then be a product of the matrices ${\bf T_1}$ and ${\bf T_2}$, i.e.,
${\bf T}={\bf T_2}{\bf T_1}$.
The amplitudes just before the end of the N-th step $A_{N}$, $B_{N}$
are related to the amplitudes $A_0$, $B_0$ at the top left side of the 
barrier by a product of N of these transfer matrices:
\begin{equation} 
{A_N\choose B_N} = {\bf T}^N\,{A_0\choose B_0}.
\label{matrixeqn}
\end{equation}
\noindent
Symmetry of the potential about $y=0$ implies that the ground state wave 
function is symmetric, i.e., $\phi_-^0(-y)=\phi_-^0(y)$. Using matching 
and integration at and around the origin where there is a negative 
delta-potential with the symmetry property of the wave function relates 
$A_0$ and $B_0$:
\begin{equation}
B_0=\frac{1+\alpha}{1-\alpha}\,A_0.
\label{symmetry}
\end{equation}
\noindent
Since we are concerned with a bound state solution, $B_N=0$. Using this 
and Eq.(\ref{symmetry}) in Eq.(\ref{matrixeqn}) enables us to get
\begin{equation}
(1-\alpha)({\bf T}^N)_{21} \, + \,(1+\alpha)({\bf T}^N)_{22}= 0,
\label{rooteqn1}
\end{equation}
\noindent
the {\it lowest positive} solution of which gives us the value of $e_-$ when 
$u_0$, $\rho$ and $N$ are specified. The expressions for the matrix 
elements $(\bf{T}^N)_{21}$ and $({\bf T}^N)_{22}$ are given in the Appendix.

\noindent
{\bf 4. Results and discussion}

\noindent
Now we consider the solution of Eq.(\ref{rooteqn1}). The result 
could be better appreciated if we compare it with the 
corresponding escape rate for the original W-potential with no barrier 
subdivision (Fig.1a). Applying the same technique as above the equation 
corresponding to Eq.(\ref{rooteqn1}) to be solved is
\begin{equation}
\alpha_0\,+\,(1-\alpha_0)\,e^{2k_0}\,=\,0
\label{rooteqn2}
\end{equation}
\noindent
where $\alpha_0= \frac{u_0}{k_0}$, 
$k_0= [u_0^2 \,-\, e_-^0]^{\frac{1}{2}}$
with $u_0=\frac{1}{2}\beta U_0$. In this case, $e_-^0=L_0^2\,E_-^0$.
The inverse of $DE_-^0$ is the time required to go from one 
minimum to the other in Fig. (1a)  and we define the corresponding 
escape rate as $DE_-^0$ for the original `W' potential. 
Then the ratio, $f_N$, of the escape rate, 
$DE_-$, over the potential with 
a certain barrier subdivision to that of escape rate, $DE_-^0$, over 
the original W-potential is given by
\begin{equation}
f_N\equiv\frac{E_-}{E_-^0}=(1+\rho)^2\, \frac{e_-}{e_-^0}.
\end{equation}
\noindent
We call this ratio, $f_N$, as the {\it enhancement factor}. It may be 
worthwhile pointing out here that we have used the first passage time 
from one minimum to the other in the original potential as a scale factor.
This is because the subdivided 
potential is rugged on the `down hill part' as well, which could give rise to
a considerably different transit time compared to  the situation if 
only `sliding down' on a smooth line were allowed.
 
There are only two parameters in our model, namely $u_0$, which 
is the total barrier height  and $\rho$, which
essentially represents the steepness of the local barriers. 
We chose $u_0$ (so as to be in the high barrier limit) holding  $\rho$ fixed
and explored the enhancement factor, $f_N$, for various 
values of barrier subdivision, $N$.  Fig. 2 shows plots of $f_N$ 
versus $N$ for three different choices of $u_0$ with fixed $\rho$(=0.8).
For this case, the enhancement factor at the optimal barrier subdivision,
$N_{op}$, increases as $u_0$ is increased reaching a value as high 
as 35 for $u_0$ = 12.0, while $N_{op}$ remains constant (here 9) suggesting 
that the steepness of the local barrier determines $N_{op}$.
In Fig. 3, we have varied $\rho$ for a fixed value of $u_0$(=9.0).
In this case, both $N_{op}$ and the enhancement factor increase
as $\rho$ is increased.( Note that in both the figures N starts at  values  
larger than 2 for higher values of $\rho$. This is due to the fact that for 
these values of $\rho$,  the enhancement factor is less 
than unity for low values of $N$ and correspond to the situation where
$U_2>U_1$.)
 
We have verified that these trends are general. Thus, {\it there 
is an optimal value of barrier subdivision, $N_{op}$, at which the escape 
rate takes a maximum value}. The existence of $N_{op}$ may be 
readily understood by a reference to the potential $V_-(x)$. The binding 
energy of this localized ground state of the individual negative energy 
delta-functions and its lowering due to mutual overlap of the neighbouring 
bound states (banding effect) are oppositely affected by $N$.  

It may be worthwhile to mention here that in addition to changing the
terrain (say steepness) of the intermediate barrier (connecting the 
initial and final states) by the subdivision, the outer barriers were 
also made to change their steepness accordingly {\it for the sake of 
simplicity} (see Figs. 1a and 1b). Because of 
this increased steepness, they become more confining than with their
original slope and, thus, give rise to an overestimation of the enhancement 
factor. We have verified
this by retaining the original slope of the outer barriers. 
However, the main features remain the same. On the other 
hand, due to the monotonic increase of the optimal enhancement factor 
with the barrier height, its value could be much larger than the ones 
considered here for barrier heights that exist in chemical and physical 
processes.

We remark that this problem can be viewed, approximately, as that of finding 
the mean first
passage time of a biased random walk [8]. However, this would imply assigning
values to the forward and the backward transition rates for the individual
sub-barriers taken {\it in isolation} as input, i.e. assuming that the 
potentials to the left and right sides of each sub-barrier are totally
confining, and then using these input values to calculate the global
escape rate for the coupled sub-barriers. We have found that while this 
gives an optimal barrier subdivision for the escape rate consistent with the
present result, the enhancement factor is considerably over-estimated by this 
random walk approach. The present SUSY-based calculation goes beyond this
uncontrolled approximation. 

It would be interesting to examine and optimize the effect of an athermal 
(possibly colored) noise (the `blow torch' of Landauer [9,10]) on one of 
the steps of our subdivided potential curve. This is under investigation.

In conclusion, we have shown that the Kramers' rate for the escape over a
given potential barrier, in the high barrier high friction limit, can be
substantially enhanced by subdividing the barrier optimally. This might 
provide an alternative scenario for certain activated processes where
the measured escape rate is substantially higher than that anticipated.

\noindent
{\bf Appendix A}

\noindent
To find the elements of the matrix ${\bf T}^N$ we decompose the transfer
matrix ${\bf T}$ as a product of three matrices; i.e.
\setcounter{equation}{0}
\renewcommand{\theequation}{A\arabic{equation}}
\begin{equation}
{\bf T}={\bf R}\,{\bf \Lambda}\,{\bf L}
\end{equation}
so that ${\bf \Lambda}$ is a diagonal matrix whose diagonal elements are the
eigenvalues of ${\bf T}$. The matrices ${\bf L}$ and ${\bf R}$ are, 
respectively, made up of the left- and right-eigenvectors of ${\bf T}$ 
such that ${\bf L}\,{\bf R}={\bf R}\,{\bf L}={\bf I}$. With this 
decomposition,
\begin{equation}
{\bf T}^N\,={\bf R}\,{\bf \Lambda}^N\,{\bf L}
\end{equation}
whose two elements of our interest, $({\bf T}^N)_{21}$ and $({\bf T}^N)_{22}$
are expressed as
\begin{equation}
({\bf T}^N)_{21}\,=\frac{T_{21}(-\lambda_{-}^N +\lambda_{+}^N)}{Q}
\end{equation}
and
\begin{equation}
({\bf T}^N)_{22}\, =
\frac{(T_{11}-T_{22})(\lambda_{-}^N- \lambda_{+}^N)+(\lambda_{-}^N+
\lambda_{+}^N)Q}{2Q}.
\end{equation}
\noindent
$\lambda_{\pm}$ are the eigenvalues of ${\bf T}$ given by
\begin{equation}
\lambda_{\pm}=\frac{1}{2}(T_{11}+T_{22}\,\pm Q)
\end{equation}
\noindent
with $T_{ij}$ as the matrix elements of ${\bf T}$ and 
$Q=[(T_{11}-T_{22})^2\,+4T_{12}T_{21}]^{\frac{1}{2}}$.

\noindent
{\bf Acknowledgements}

\noindent
We would like to thank Dr. A.M. Jayanavar for bringing ref. 7 to our
attention. One of us (M.B.) would like to thank The International Program
in Physical Sciences (Uppsala University, Sweden) for financial support
for this work. M.B. would also like to acknowledge Dr. M. P. Joy for his
technical advice on computation.

\centerline{\bf Figure Captions}

Figure 1: (a) The model potential.
          (b) Plot of the subdivided model potential: $U(x)$ versus $x$,
              when $N=3$. Note the change in the slope of the left- and
              right-confining walls from that of (a).
          (c) Plot of $V_-(x)$ versus $x$ (not to scale). 
          (d) A figure showing a step of the subdivided potential found
              between the intervals $x_n$ and $x_{n+1}$.

Figure 2: Plots of $f_N$ versus $N$ for three different values of $u_0$ 
          (i.e. 6.0, 9.0, 12.0) with fixed $\rho(=0.8)$. Note that all the 
          optimal values occur at $N=9$. 

Figure 3: Plots of $f_N$ versus $N$ for four different values of $\rho$ 
          (i.e. 0.4, 0.6, 0.8, 0.9) with fixed $u_0(=9.0)$. 
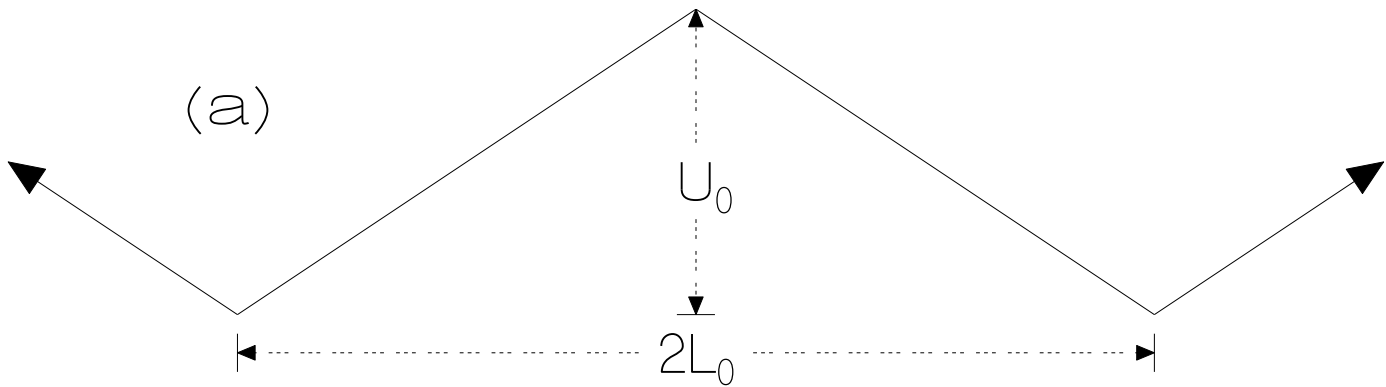
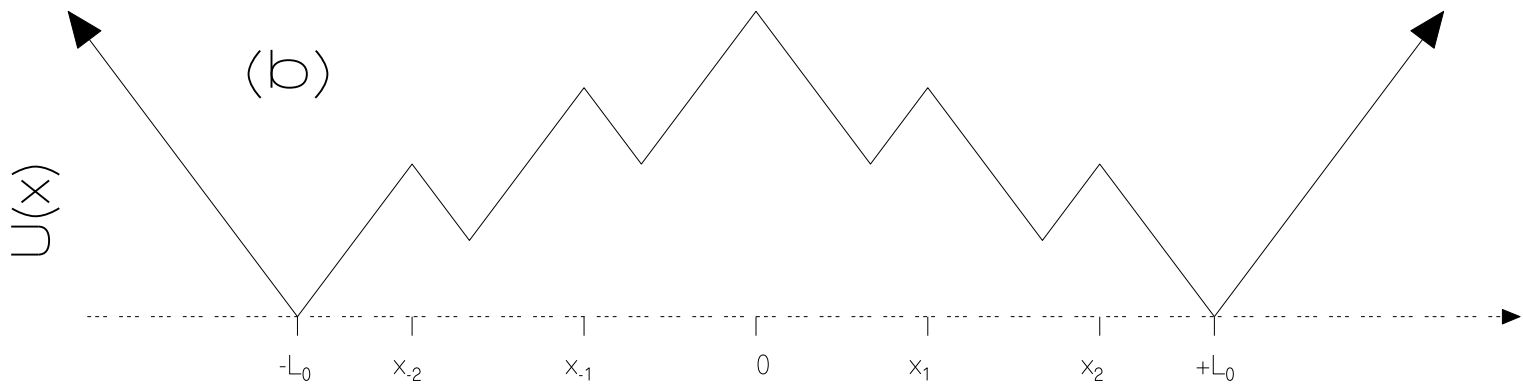
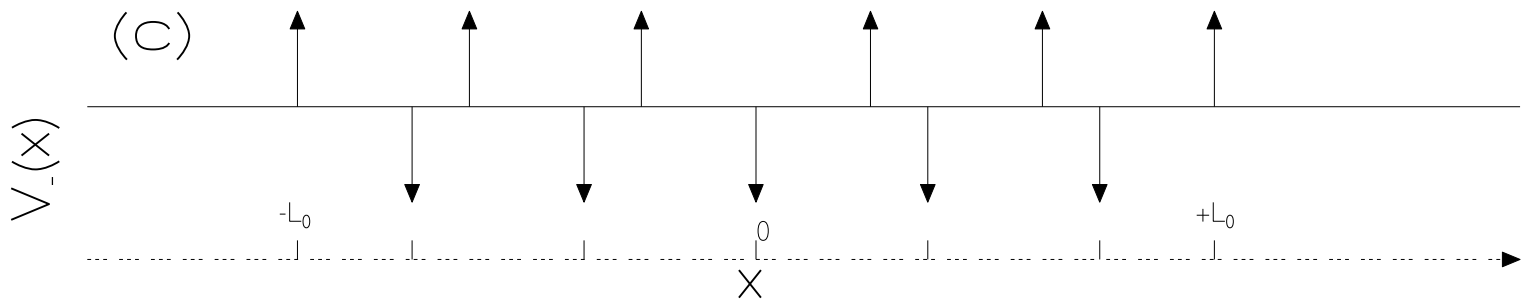
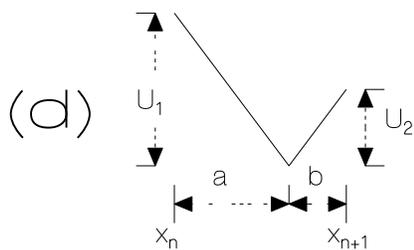
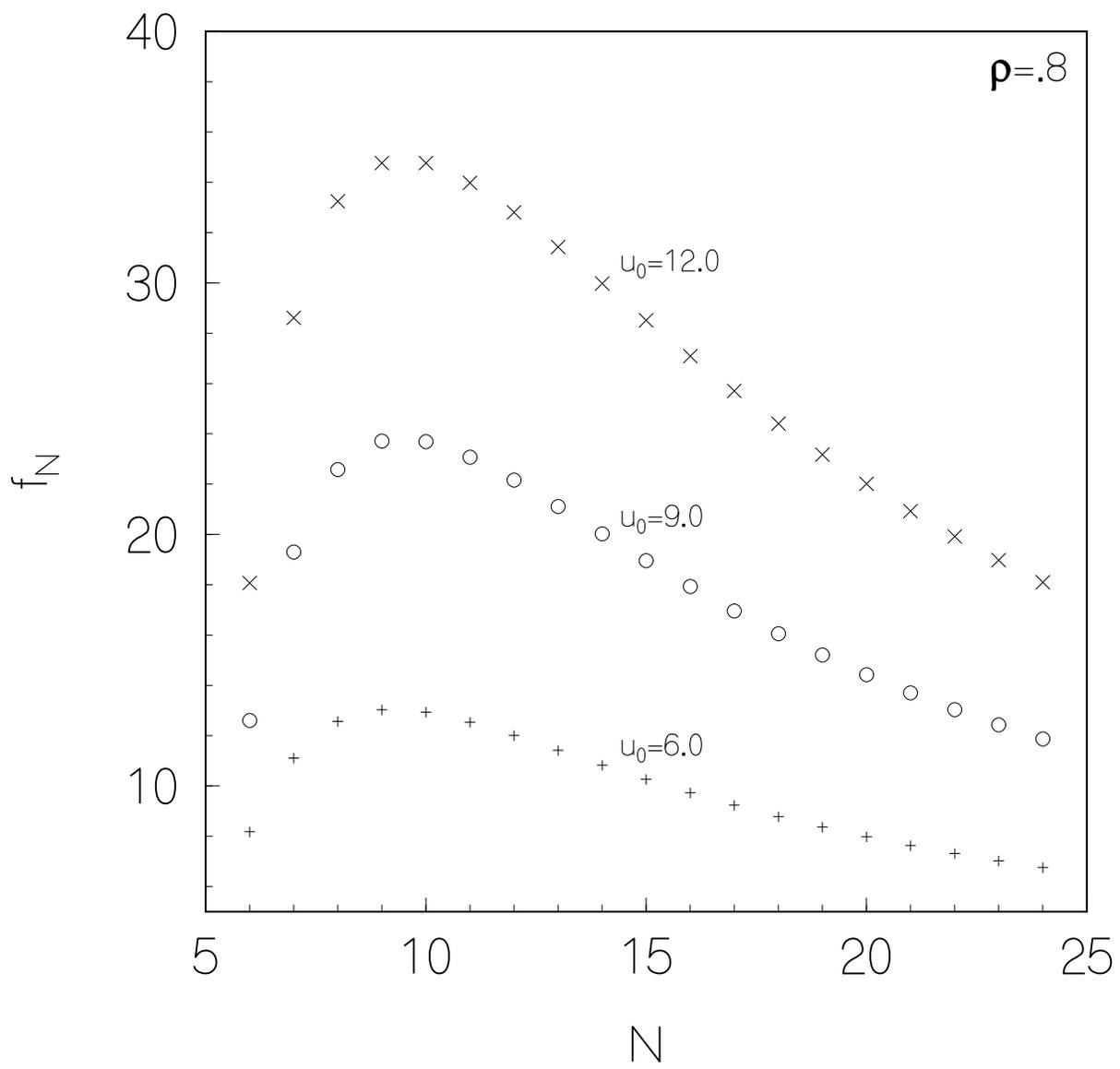
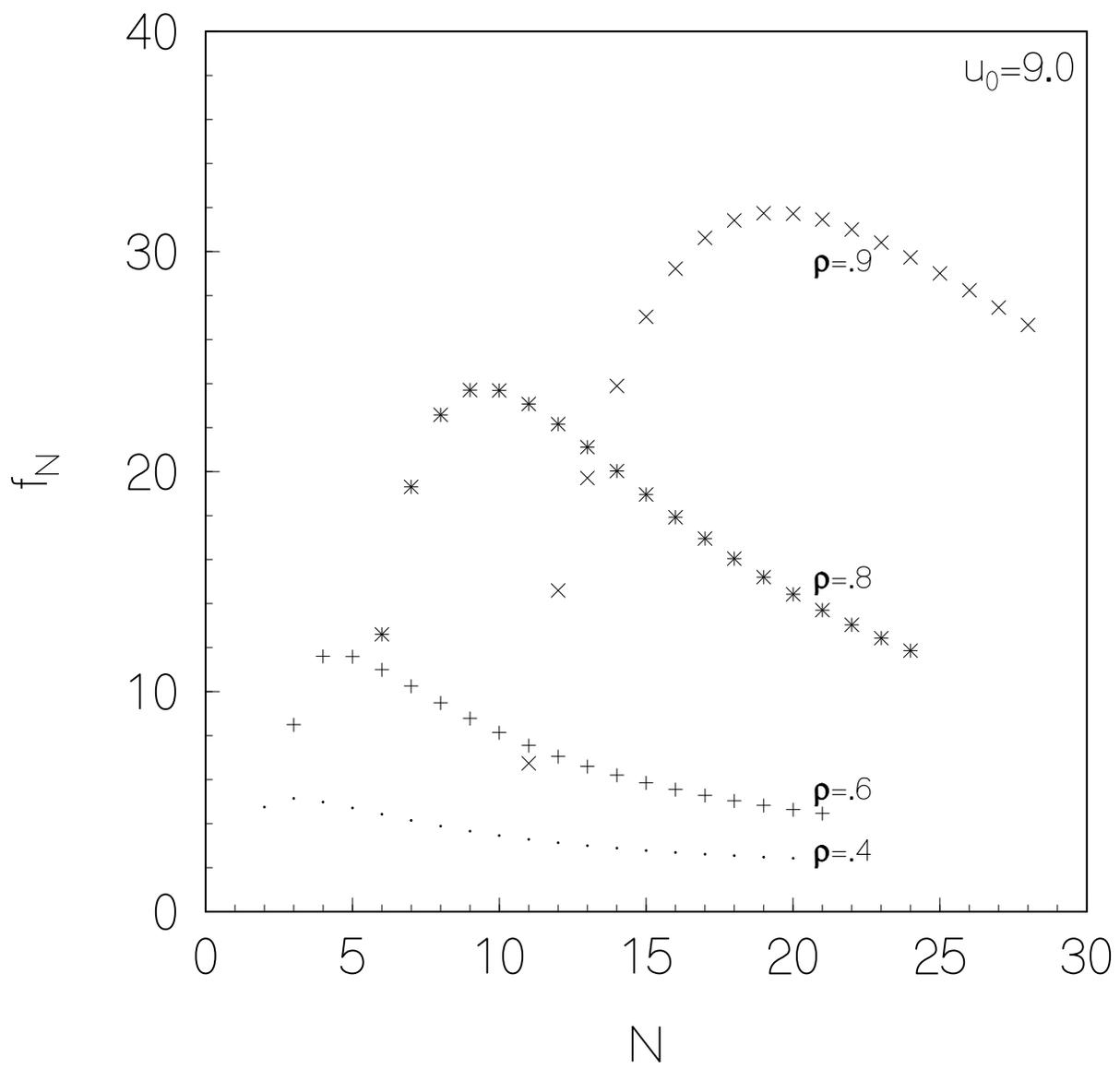
\end{document}